\documentclass[conference]{IEEEtran}
\IEEEoverridecommandlockouts
\usepackage{cite}
\usepackage{amsmath,amssymb,amsfonts}
\usepackage{algorithmic}
\usepackage{graphicx}
\usepackage{textcomp}
\usepackage{xcolor}

\usepackage{balance}

\usepackage{amsmath}
\usepackage{pifont}
\newcommand{\cmark}{\ding{51}}%
\newcommand{\xmark}{\ding{53}}%
\newcommand{\cxmark}{\ding{55}}%

\usepackage{threeparttable}

\usepackage{stfloats}
\usepackage{hyperref}
\usepackage{booktabs}
\usepackage{multicol,multirow}
\usepackage[table]{xcolor}
\usepackage{hhline}
\definecolor{TBLHeader}{gray}{0.85}
\definecolor{TBLRow}{gray}{0.95}

\definecolor{TBBRow6}{HTML}{62A8D2}
\definecolor{TBBRow5}{HTML}{79b9da}

\definecolor{TBBRow4}{HTML}{92c9e2}
\definecolor{TBBRow3}{HTML}{abdaeb}
\definecolor{TBBRow2}{HTML}{c5eaf4}
\definecolor{TBBRow1}{HTML}{e0faff}

\definecolor{AdRed}{rgb}{0.831, 0.047, 0.047}
\usepackage{graphicx}

\def\BibTeX{{\rm B\kern-.05em{\sc i\kern-.025em b}\kern-.08em
    T\kern-.1667em\lower.7ex\hbox{E}\kern-.125emX}}
\makeatletter
\let\old@ps@headings\ps@headings
\let\old@ps@IEEEtitlepagestyle\ps@IEEEtitlepagestyle
\def\confheader#1{%

	\def\ps@IEEEtitlepagestyle{%
		\old@ps@IEEEtitlepagestyle%
		\def\@oddhead{\parbox[b]{9cm}{\raggedright \@IEEEheaderstyle #1}}%
		\def\@evenhead{\strut\hfill#1\hfill\strut}%
	}%
	\ps@headings%
}
\makeatother

\confheader{%
	\textbf{\textit{2025 IEEE International Conference on Energy Technologies for Future Grids \\                                                                                                     
			Wollongong, Australia, December 7-11, 2025}}
}

\begin{document}

\title{A new measure to quantify the similarity of load profile time-series\\
\thanks{This project is supported by the Australian Government Research Training Program (RTP) through the University of Adelaide, and a supplementary scholarship provided by Watts A/S, Denmark.}
}

\author{\IEEEauthorblockN{Rui Yuan}
\IEEEauthorblockA{\textit{School of Elec. \& Mech. Eng.} \\
\textit{The University of Adelaide}\\
Adelaide, Australia \\
r.yuan@adelaide.edu.au}
\and
\IEEEauthorblockN{Hossein Ranjbar}
\IEEEauthorblockA{\textit{School of Elec. \& Mech. Eng.} \\
\textit{The University of Adelaide}\\
Adelaide, Australia \\
hossein.ranjbar@adelaide.edu.au}
\and
\IEEEauthorblockN{S. Ali Pourmousavi}
\IEEEauthorblockA{\textit{School of Elec. \& Mech. Eng.} \\
\textit{The University of Adelaide}\\
Adelaide, Australia \\
a.pourm@adelaide.edu.au}
\and
\IEEEauthorblockN{Wen L. Soong}
\IEEEauthorblockA{\textit{School of Elec. \& Mech. Eng.} \\
\textit{The University of Adelaide}\\
Adelaide, Australia \\
wen.soong@adelaide.edu.au}

\and
\IEEEauthorblockN{Andrew J. Black}
\IEEEauthorblockA{\textit{School of Comp. \& Math. Sciences} \\
\textit{The University of Adelaide}\\
Adelaide, Australia \\
andrew.black@adelaide.edu.au}
\and
\IEEEauthorblockN{Jon A. R. Liisberg}
\IEEEauthorblockA{\textit{Research \& Dep. of Data Science} \\
\textit{Watts A/S}\\
Zealand, Denmark \\
jon.liisberg@watts.dk}
\and
\IEEEauthorblockN{Julian Lemos-Vinasco}
\IEEEauthorblockA{\textit{Research \& Dep. of Data Science} \\
\textit{Watts A/S}\\
Zealand, Denmark \\
julian.lemos@watts.dk}
}

\IEEEoverridecommandlockouts
\IEEEpubid{\makebox[\columnwidth]{979-8-3315-7640-0/25/\$31.00~\copyright2025 IEEE \hfill} \hspace{\columnsep}\makebox[\columnwidth]{ }}
\maketitle
\IEEEpubidadjcol

\begin{abstract}
This paper proposes a novel time-series similarity metric, called Flexibility Distance (FD), to quantify the effort required to reshape one energy profile into another by jointly considering amplitude modifications and temporal shifts. Unlike conventional measures, such as Euclidean Distance (ED) and Dynamic Time Warping (DTW), FD is specifically designed to capture key aspects of load flexibility, including peak shaving, load shifting, and adaptive consumption. This enables more effective demand-side management, improves grid efficiency, and facilitates the integration of renewable energy. The proposed metric is tested in a simulation framework where consumer load profiles are rescheduled to match ideal flexibility targets. Comparative analyses show that FD outperforms ED and DTW, which do not adequately account for the temporal adaptability of energy consumption. The simulation results demonstrate that FD more accurately evaluates demand flexibility, yielding closer alignment with the ideal rescheduling profiles.
\end{abstract}
\vspace{1em}
\begin{IEEEkeywords}
Energy analysis, continuous demand response, demand side management, flexibility distance.
\end{IEEEkeywords}

\section{Introduction}
\label{sec: intro}
Traditional power systems are evolving into smarter, more efficient, and more sustainable electric networks, aka smart grids. With digital devices, smart sensors, higher and cheaper edge computing power, and sophisticated control algorithms, various smart grid applications are expected to bring numerous benefits to end users and network operators and enable smooth and cost-effective decarbonisation of energy systems. As such, governments have made huge investments in developing enabling technologies, and numerous smart grid applications have already been developed, e.g., non-intrusive load modelling (NILM) and continuous demand response (CDR), or even deployed on a large scale, e.g., home energy management systems through grid-interactive efficient buildings initiatives~\cite{IEA2023}. At their core, these applications assess the similarity between two or more time series~\cite{GRANELL2015, YUAN2023105588}. As a result, similarity metrics, especially distance-based methods, play a critical role in the implementation, performance, and fairness of these algorithms. This is because the magnitude of energy usage is critical, whereas correlation-based methods, e.g., cosine similarity or Pearson correlation, focus on pattern similarity regardless of absolute magnitude.

In the literature, Euclidean Distance (ED) and Dynamic Time Warping (DTW) are the two most commonly used similarity measures for time series comparison in energy applications~\cite{Lin2012, Fang2018, YUAN2023105588}. Several studies have also used other basic spatial metrics, such as the L1 norm, Manhattan, and Chebyshev distances, but their performance is often reported to be comparable to ED~\cite{Nichiforov2021}. While variants of DTW have been developed for specific domains~\cite{YUAN2023105588, shaw2025forestproximitiestimeseries}, their ability to meaningfully quantify flexibility in energy use remains limited. ED, for instance, only captures differences in amplitude, completely ignoring temporal shifts in consumption. DTW improves on this by allowing elastic alignment of time series, making it suitable for tasks like load signature detection~\cite{Elafoudi2014}. However, it treats all temporal shifts uniformly, failing to distinguish between short- and long-term load rescheduling, which is an important consideration for energy flexibility assessment.

Moreover, the concept of flexibility, referring to the ability to adjust or reschedule energy consumption over time, has gained increasing attention in the CDR literature~\cite{Wen2023, Li2024, Goldenberg2018, Gerards2017, Varenhorst2024}. Flexibility has been defined in various ways: as the amount of energy that can be increased or decreased within a specified time window~\cite{flex2012, Wen2023, Goldenberg2018}, or as a binary indicator reflecting whether specific consumption adjustments are delivered~\cite{Gerards2017, Varenhorst2024}, often using causality-based methods to assess the contribution of individual appliances. However, these approaches typically rely on ED or its conceptual variants, which implicitly assume that the effort required to shift a load is the same across time, which is a simplification that fails to capture the temporal complexity and behavioural dynamics of real energy use.

To address these limitations, this paper introduces a novel similarity measure, termed Flexibility Distance (FD), specifically designed to quantify the time series distance, and hence flexibility measurement in energy consumption profiles. FD jointly considers both amplitude variations and temporal shifts, capturing behaviours such as load increases, decreases, rescheduling, and sequence reordering. By evaluating the real effort required to transform one time series into another, FD offers a more comprehensive and interpretable metric for use in energy data analytics. Although developed with electricity consumption in mind, FD is applicable to more general tasks of \emph{range queries} and \emph{nearest neighbour queries} in time series mining. In summary, the main contributions of this paper are as follows:

\begin{itemize}
    \item Proposing the FD measure as a new time series similarity metric that explicitly captures both amplitude and temporal differences.
    
    \item Demonstrating, through a comprehensive simulation study with multiple scenarios, that FD outperforms traditional metrics (ED and DTW) in quantifying flexibility and aligning with ideal rescheduling profiles.
    
    \item Illustrating that FD provides a more meaningful basis for evaluating flexibility in CDR applications, with implications for improved grid operation and renewable energy integration.
\end{itemize}

The remainder of this paper is structured as follows. Section~\ref{sec:definition} introduces the problem and highlights the limitations of existing similarity measures. Section~\ref{sec:method} outlines the proposed methodology, developed in line with the criteria for an ideal similarity metric, and compares it against ED and DTW. Section~\ref{sec:experiment} presents simulation studies and provides a detailed analysis of the results. Finally, Section~\ref{sec:conclusions} concludes the paper.


\section{Problem Definition}
\label{sec:definition}

Consider two time series of length $m$, $\mathbf{X} = \{X_1,...X_m\}$ and $\mathbf{Y} = \{Y_1,...Y_m\}$. The similarity of the two time series defined by ED is calculated by accumulating the point value differences at the same timestamp:
\begin{align}
    \text{ED}(\mathbf{X}, \mathbf{Y})=\sqrt{\sum^m_{i=1}(X_i-Y_i)^2}\label{eq:Ed}
\end{align}

On the other hand, DTW finds the minimum warping path and its associated distance of the two series~\cite{Keogh2005}:
\begin{equation}
    \text{DTW}(\mathbf{X}, \mathbf{Y}) = \sqrt{\Theta_{X_{m},Y_{m}}},\label{eq:DTW}
\end{equation}
\noindent where $X_m$ and $Y_m$ are the $m$\textsuperscript{th} point in $\mathbf{X}$ and $\mathbf{Y}$, $\Theta_{X_{m},Y_{m}}$ is the cumulative distance of $\mathbf{X}$ and $\mathbf{Y}$ from $1$ to $m$. The distance between $i$\textsuperscript{th} point in $\mathbf{X}$ and $j$\textsuperscript{th} point in $\mathbf{Y}$ can be calculated as:
\begin{align}
    \!\!\Theta_{X_i,Y_j} \!\! = \! (X_i\! - \! Y_j)^2 \!\! + \! \min\{\Theta_{X_{i\! - \! 1},Y_{j\! -\! 1}},\! \Theta_{X_{i\! - \! 1}, Y_j},\! \Theta_{X_i,Y_{j\! - \! 1}}\!\}.\label{eq:DTW-2}
\end{align}

\begin{figure}[!tb]
    \centering
    \includegraphics[width=.5 \textwidth]{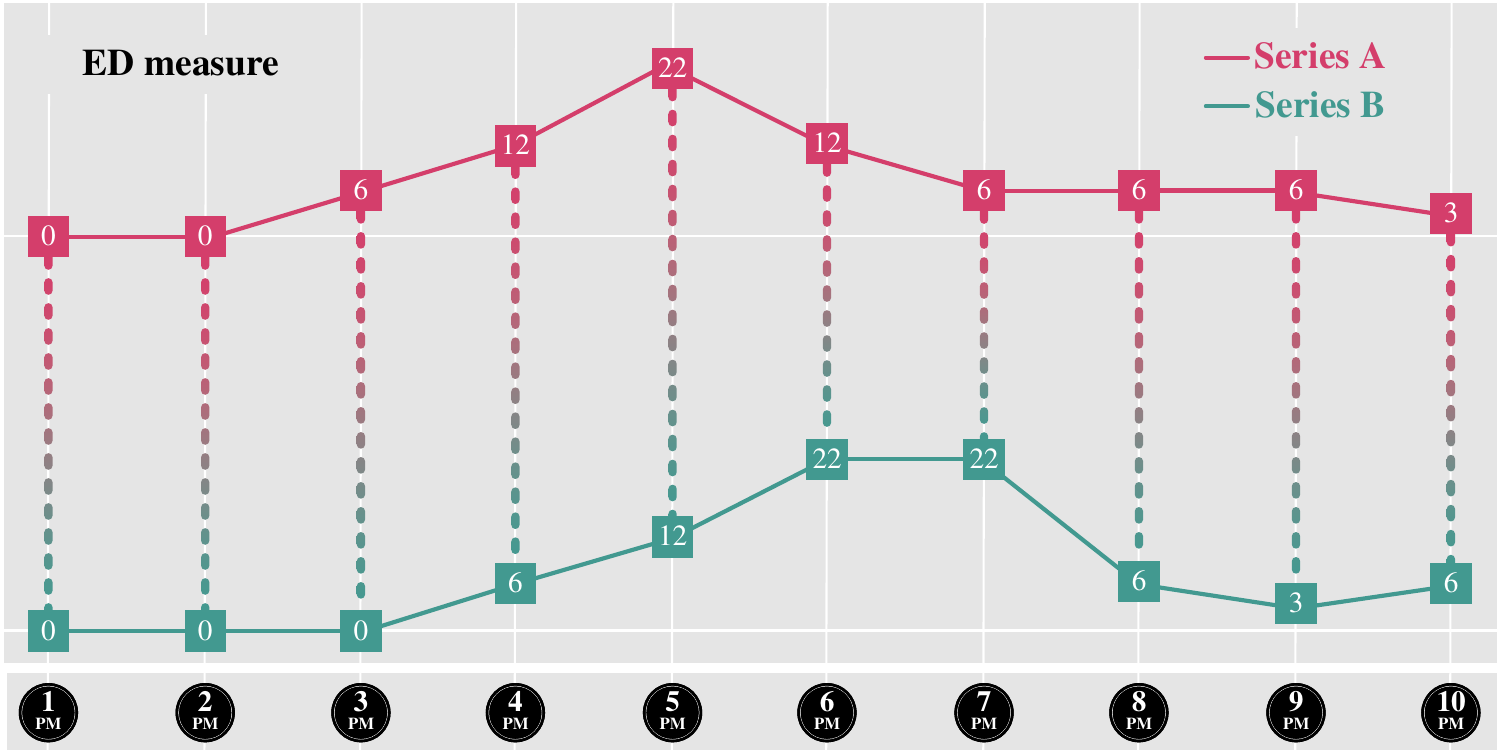}\vspace{0.2cm}
    \includegraphics[width=.5 \textwidth]{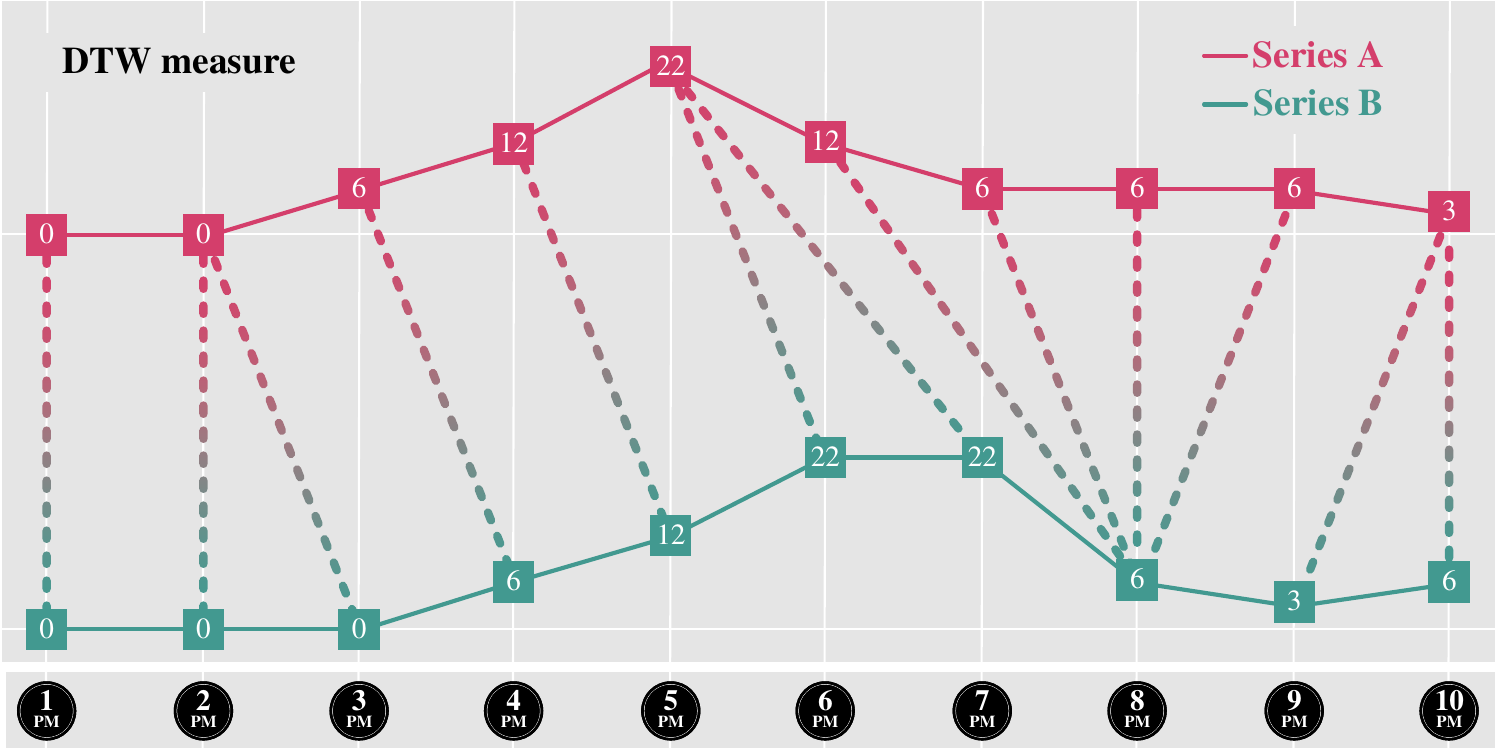}
    \caption{Distance measurement comparison with ED (top) and DTW (bottom) for a hypothetical electric vehicle (EV) charging time series $\mathbf{A}$ (red) and time series $\mathbf{B}$ (green). The dashed line represents the selected route $r$ to accumulate the point difference between two time series by ED and DTW}
    \label{fig:EDDTW}\vspace{-10pt}
\end{figure}

More specifically, ED only measures the amount of energy change at each interval, whereas DTW considers two time series similar if they have similar patterns, even if these occur at different times. This difference is shown in an illustrative example in Fig.~\ref{fig:EDDTW}, where we compare two data sets of EV charging between 1 pm and 10 pm. ED evaluates electricity usage at identical intervals, ignoring any shifts in time. However, DTW accounts for temporal shifts by matching the value at a specific point in series $\mathbf{A}$ with the values at the preceding, same, and subsequent intervals in series $\mathbf{B}$. For example, the electricity consumption at 3 pm in series $\mathbf{A}$ will be compared with the values at 4 pm in series $\mathbf{B}$, resulting in a zero difference. However, from a power system engineering point of view, these two instances in time series $\mathbf{A}$ and $\mathbf{B}$ are not identical because it requires a certain amount of effort, incentive, or compromise in comfort to shift the load to an hour later. This capability to measure similarity in the presence of temporal variations, incorporating additional domain-specific factors, is hereafter termed \emph{temporal sensitivity}. 

Additionally, DTW finds the EV charging at 5 pm in time series $\mathbf{A}$ identical to the values from noon to 6 pm and 7 pm in time series $\mathbf{B}$, while these two patterns are not similar. In time series $\mathbf{A}$, 22 kWh is consumed in one hour, whereas in time series $\mathbf{B}$, we see a shift in EV charging by one hour and an extension of the charging for the next hour. As a result, these two patterns show different levels of energy consumption, starting points, and duration, which makes them dissimilar. This characteristic to which DTW is insensitive is hereafter referred to as \emph{temporal uniqueness}.

Thus, given two energy profiles $\mathbf{X}$ and $\mathbf{Y}$, a distance matrix with elements $d_{i,j} = (X_i-Y_j)^2$ contains the point-by-point distance between $X_i$ and $Y_j$. A route selection matrix with elements $r_{i,j} \in \{0, 1\}$, specifies the path to compute the similarity of $\mathbf{X}$ and $\mathbf{Y}$ by accumulating a subset of point-by-point distances in $d_{i,j}$. Hence, an ideal measure that can compare the electricity profiles and quantify the flexibility, $M(\mathbf{X}, \mathbf{Y}) = \sum_{i=1}^m \sum_{j=1}^m r_{i,j}\cdot d_{i, j}$ should meet the following seven requirements:

\begin{enumerate}
    \item Non-negativity: $M(\mathbf{X}, \mathbf{Y}) \geq 0$
    \item Identity: $M(\mathbf{X}, \mathbf{Y}) = 0$, if and only if $\mathbf{X} = \mathbf{Y}$
    \item Symmetry: $M(\mathbf{X}, \mathbf{Y}) = M(\mathbf{Y}, \mathbf{X})$
    \item Triangle inequality: $M(\mathbf{X}, \mathbf{Y}) \leq M(\mathbf{X}, \mathbf{Z}) + M(\mathbf{Z}, \mathbf{Y})$ for any time series $\mathbf{X}, \mathbf{Y}, \mathbf{Z}$
    \item Temporal uniqueness: $\sum_{i=1}^m r_{ij} = 1\;\forall \;j \in \{1,\dots, m\}$
    \item Temporal sensitivity: $d_{i, j}\neq 0$, if $i\neq j$
    \item Optimal match:  $M(\mathbf{X}, \mathbf{Y}) = \displaystyle \min_{r \in \mathcal{R}} \{\sum_{i=1}^m \sum_{j=1}^m r_{i,j}d_{i,j}$\}
\end{enumerate}

Upon comparing ED and DTW with the ideal requirements outlined above, it becomes evident that ED fails to satisfy Requirements~5 to~7. DTW, on the other hand, does not meet Requirements~2, 5, and~6, and only partially satisfies Requirement~3, depending on whether a symmetric step pattern is used in the route selection matrix~\cite{Keogh2005}.

\begin{figure}[!tb]
    \centering
    \includegraphics[width=.4 \textwidth]{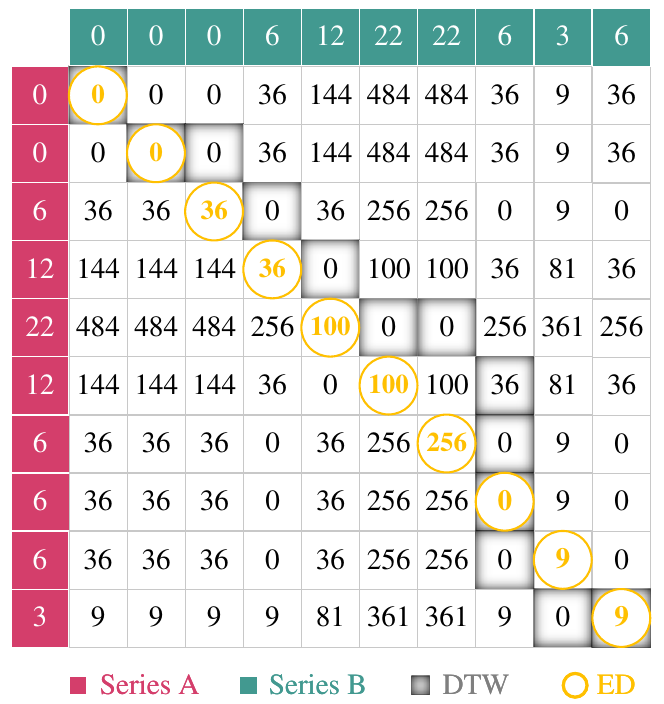}
    \caption{Distance measurement comparison using sample time series $\mathbf{A}$ and $\mathbf{B}$ from Fig.~\ref{fig:EDDTW}. The distance matrix with $d_{i,j}$ is calculated by pointwise comparison. The routes indicate the accumulated distance by ED (yellow circles) and DTW (shaded squares)}
    \label{fig:med-EDDTW}\vspace{-10pt}
\end{figure}

A detailed comparison between the two methods is illustrated in Fig.~\ref{fig:med-EDDTW}, using two sample time series shown in Fig.~\ref{fig:EDDTW}. The distance matrix in Fig.~\ref{fig:med-EDDTW} is constructed using the point-wise squared difference, $d_{i,j} = (A_i - B_j)^2$. The selected alignment route is represented as a binary matrix, where $r_{i,j} = 1$ indicates that element $i$ from series $\mathbf{A}$ is aligned with element $j$ from series $\mathbf{B}$. For visual clarity, the alignment path is marked using yellow circles for ED and shaded squares for DTW.

The ED and DTW distance values are then computed using~\eqref{eq:Ed} and~\eqref{eq:DTW}, respectively. As shown, ED only accounts for the pairwise differences between the corresponding values at the same time intervals, that is, along the diagonal of the distance matrix. In contrast, DTW captures dissimilarities in temporal patterns by computing an optimal warping path that minimises the cumulative distance between aligned points. This allows DTW to recognise ``stretched'' or ``shifted'' patterns as similar by aligning one point in a time series with multiple points in another.

However, DTW's flexibility in warping leads to a fundamental issue in this application domain: it allows an interval in series $\mathbf{A}$ to be matched with multiple intervals in series $\mathbf{B}$, which contradicts physical and operational constraints in energy rescheduling scenarios. As a result, DTW fails to satisfy Requirement~5, which mandates temporal alignment constraints that reflect realistic load-shifting behaviour.

\section{The Proposed Methodology}
\label{sec:method}

In practice, electricity consumption patterns are influenced by a variety of behavioural and contextual factors, leading to shifts, stretches, and rearrangements in temporal usage. For instance, an EV might be charged during the day instead of overnight, cooking may take an hour longer than usual, or television may be watched before dinner instead of after. These temporal variations must be taken into account in any similarity measure, as they directly affect consumer welfare, an aspect captured in Requirement 6, which is not satisfied by ED or DTW.

Moreover, transforming one usage pattern into another typically allows multiple matching paths with varying associated costs. To achieve optimal matching, the solution with the minimum total cost should be selected, as described in Requirement 7. This criterion is unmet by the ED and is only partially addressed by the DTW, which identifies a minimum-cost path, but does so through a constrained warping mechanism.

To address these limitations, we introduce a new distance metric, termed \emph{Flexibility Distance} (FD), designed to more effectively quantify the similarity between the electricity usage profiles. FD operates in two phases. The first phase constructs a cost matrix that simultaneously accounts for both amplitude and temporal misalignment. Each element $D_{i,j}$ in this matrix represents the effort required to reshape a consumption value $X_i$ into $Y_j$, incorporating both the magnitude and the timing of the adjustment.
\begin{align}
    D_{i,j} = |X_{i} - Y_{j}|\cdot P_{i,j}+ |i - j|\cdot T_{i,j}
    \label{eq:fd-2}
\end{align}
\noindent where $P_{i,j}$ and $T_{i,j}$ are the amplitude and temporal weights, respectively. For illustration purposes, we use a constant number $1$, as the amplitude weight, and a max-min scaler of $X_i$ as the temporal weight to make the amplitude and temporal effects similar for simplicity. Hence, $D_{i,j}$ can be calculated using ~\eqref{eq:fd-3}. In real-world applications, different weights can be set according to the needs.
\begin{align}
    D_{i,j} \!= \!|X_{i} - Y_{j}|\!+\! |i - j|\! \cdot \!\frac{\max(X,Y) - \min(X,Y)}{m}\label{eq:fd-3}
\end{align}

To represent the possible route that fulfils \emph{temporal uniqueness} of requirement 5, we define $r_{i,j}$ as binary matrix elements, where each row and column have only one non-zero element, i.e., a route selection matrix, subject to:
\begin{subequations}
\begin{align}
    r_{i,j} \in \{0,1\}\; \forall \;i,j \in \{1,2,\dots, m\}\\
    \sum_{j=1}^m r_{i,j} = 1\;\forall \;i \in \{1,2,\dots, m\}\\
    \sum_{i=1}^m r_{i,j} = 1\;\forall \;j \in \{1,2,\dots, m\}
     \label{eq:fd-1} 
\end{align}
\end{subequations}

In the second phase, we select the optimal route to fulfil Requirement 7. The FD metric can then be calculated as follows:
\begin{equation}
    \text{FD}(\mathbf{X},\mathbf{Y}) = \min\limits_{r_{i,j}}\{\sum_{i=1}^m \sum_{j=1}^m r_{i,j} \cdot D_{i, j}\} \label{eq:FD-1}
\end{equation}

\begin{figure}[!tb]
    \centering
    \includegraphics[width=.4 \textwidth]{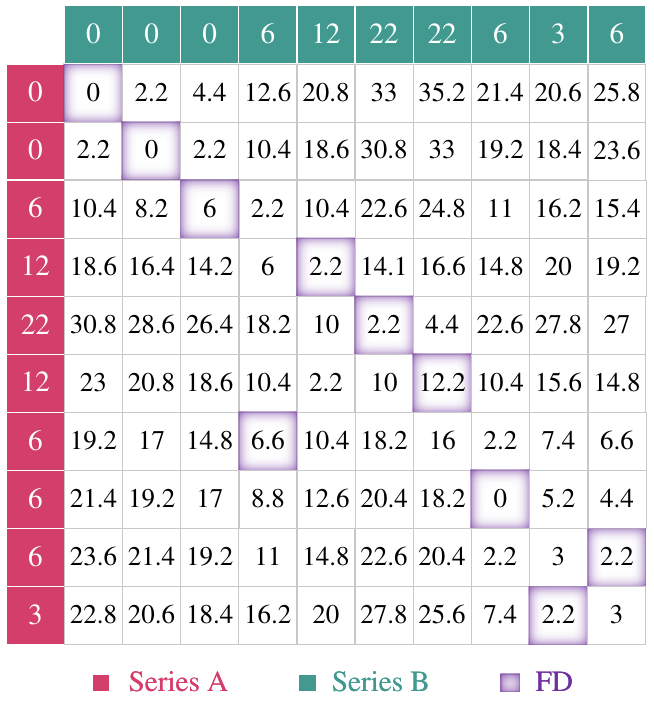}
    \caption{Distance measurement comparison using two time series $\mathbf{A}$ and $\mathbf{B}$  from Fig.~\ref{fig:EDDTW}. The proposed distance measure (shaded squares) with an improved distance matrix with elements $C_{i,j}$ calculated using~\eqref{eq:fd-3}}
    \label{fig:med-FD}\vspace{-10pt}
\end{figure}

The main idea in the proposed distance measure is to see how one time series can be reshaped into another time series at a minimum cost. In contrast to the pointwise distance matrix of ED and DTW shown in Fig.~\ref{fig:med-EDDTW}, our proposed method considers the cost of both the amplitude and temporal changes in the first phase, i.e.,~\eqref{eq:fd-2} and ~\eqref{eq:fd-3}. For example, if we move the load consumption from time $i$ in the first time series to a different time $j$ in the other time series, the ED and DTW cannot distinguish the difference, that is, $D_{3,4}= 0$ when $X_3 = Y_4$ and $3\neq 4$. However, the FD metric can distinguish this difference, as shown in Fig.~\ref{fig:med-FD}. The second phase aims to minimise the cost of reshaping the time series $\mathbf{X}$ into $\mathbf{Y}$ by~\eqref{eq:FD-1}, which is a linear sum assignment problem (LSAP). The LSAP problem can be solved by the Kuhn-Munkres Hungarian Algorithm with $O(n^3)$ time and $O(n^2)$ space~\cite{Bougleux2020}. This process involves considering all possibilities and combinations in increment, decrease, shift, and temporal sequence exchange.

Table~\ref{tab: feature-comparison} summarises the capability of ED, DTW and FD to meet the requirements listed in Section~\ref{sec:method}. It is important to note that, under certain circumstances, if one of the two time series is flat, i.e., $X_i= c$ in~\eqref{eq:fd-2}, the cost of reshaping using FD will be equivalent to the sum of diagonal elements in the distance matrix. This is similar to ED but with a different magnitude, as the minimum cost is achieved when $i=j \;\; \forall \; i,j \in \{1,2,\dots, m\} $ in~\eqref{eq:FD-1}. 

Furthermore, an important by-product of the proposed distance metric is that it finds the shortest path in Fig.~\ref{fig:med-FD}, which provides an optimal solution to reshaping time series $\mathbf{X}$ into time series $\mathbf{Y}$. Hence, FD as a metric provides a quantitative distance that presents both the cost of reshaping the electricity time series and the strategy for optimal reshaping. To this end, FD could be used in many smart grid applications, e.g., to design better home energy management systems and demand response programs. 
\begin{table}[!tb]
    \centering
    \scriptsize
    \begin{threeparttable}
    \caption{Feature comparison of different distance measures}
    \label{tab: feature-comparison}
    \renewcommand{\arraystretch}{1.5}
    \begin{tabular}{ccccc}
        \hline
        \rowcolor{TBLHeader}Requirements & & ED& DTW & FD\\
        \cline{1-1}\cline{3-5}
         Non-negativity & & \cmark & \cmark& \cmark\\
         \cline{1-1}\cline{3-5}
         Identity & & \cmark &\xmark& \cmark  \\
         \cline{1-1}\cline{3-5}
         Symmetry & & \cmark &\cxmark& \cmark \\
         \cline{1-1}\cline{3-5}
         Triangle inequality & & \cmark &\xmark& \cmark \\
         \cline{1-1}\cline{3-5}
         Temporal uniqueness & & \cmark &\xmark& \cmark \\
         \cline{1-1}\cline{3-5}
         Temporal sensitivity & & \xmark &\xmark& \cmark \\
         \cline{1-1}\cline{3-5}
         Minimum cost & & \xmark &\cxmark& \cmark \\
         \hline
    \end{tabular}
    \begin{tablenotes}
      \item \small\cmark: accomplished requirements, \xmark: unfulfilled requirements, \cxmark: requirements partially fulfilled.
    \end{tablenotes}
    \end{threeparttable}\vspace{-10pt}
\end{table}

\section{Simulation Results}
\label{sec:experiment}

\begin{figure}[!bt]
    \centering
    \includegraphics[width=.5 \textwidth]{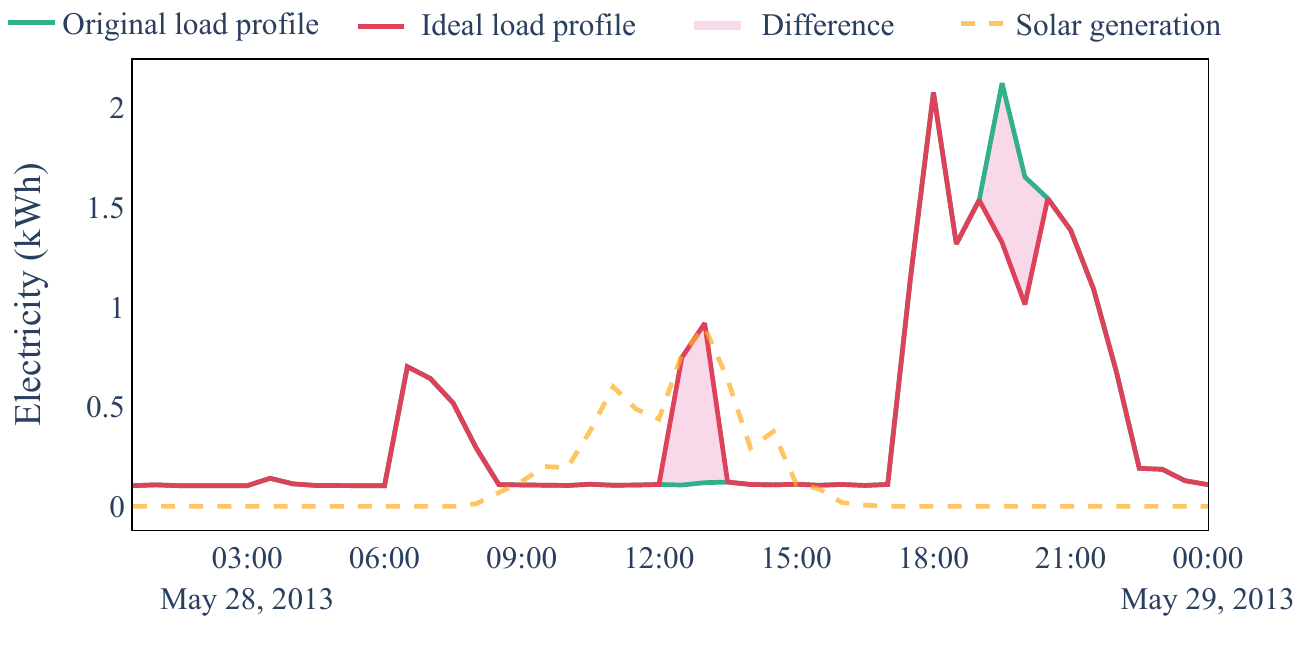}
    \caption{The original load profile (solid green line), the desired (ideal) load curve (solid red line) and solar generation (orange dashed line).}
    \label{fig: flex-comparison_A}
\end{figure}

To further evaluate the effectiveness of the proposed method against ED and DTW, we conducted a simulation study to compare different similarity measures in a CDR program using real-world data from a residential consumer with a rooftop PV system in Sydney, Australia, obtained from the SolarHome dataset~\cite{datasourcewebsite}. A specific day, i.e. 28 May 2013, was selected for analysis~\cite{datasourcewebsite}, during which both solar generation and the evening peak demand were significant, as illustrated in Fig.~\ref{fig: flex-comparison_A}. The objective in this scenario is to reshape the original load profile (green curve) into an \emph{ideal load profile} (red curve) that minimises the electricity bill by shifting consumption from the peak evening hour at 7:30 pm to the off peak noon period, when solar generation is abundant.

\begin{figure}[!bt]
    \centering
    \includegraphics[width=.5 \textwidth]{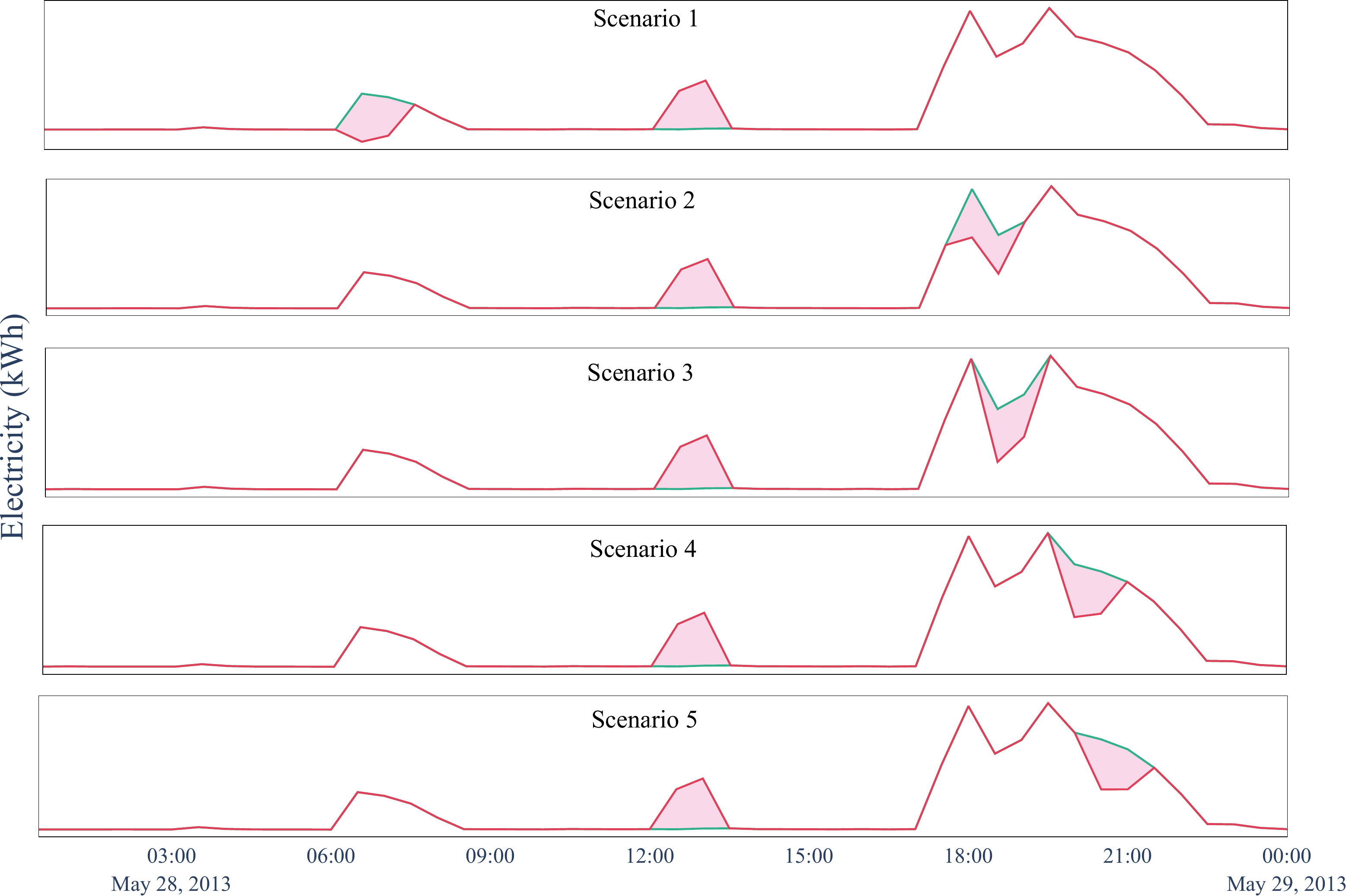}
    \caption{The proposed load rescheduling in each scenario (solid red line) in comparison with the original load profile (solid blue line) with the difference (pink shaded region).}
    \label{fig: flex-comparison_B}\vspace{-5pt}
\end{figure}

To demonstrate the utility of FD in quantifying actual flexibility, particularly in cases where the realised profile only partially follows the ideal, we generated five hypothetical rescheduling scenarios (labelled \#1--5), each involving load shifts at different time intervals, as shown in Fig.~\ref{fig: flex-comparison_B}. The amount of shifted load in each scenario corresponds to the available solar generation, as depicted in Fig.~\ref{fig: flex-comparison_A}. Notably, Scenario \#1 represents a special case where the peak load is shifted from morning to noon. While this aligns with solar availability, it does not contribute to reducing the evening peak and, in fact, introduces a new morning peak, one that is temporally distant from the intended evening shift.

To quantify the effectiveness of each scenario in approximating the ideal profile, we calculated the distance between each scenario and the ideal load profile using the three similarity metrics: ED, DTW, and the proposed FD. The results are summarised in Table~\ref{tab: distance-comparison}, with FD values scaled for comparison. The first row shows the distance between the original load profile $\mathbf{O}$ and the ideal profile $\mathbf{E}$, denoted as $M(\mathbf{O}, \mathbf{E})$, which represents the baseline effort required to fully reschedule the original load. Lower positive distances for a given scenario indicate a closer alignment with the ideal profile and thus a more effective rescheduling outcome.

It can be seen in Table~\ref{tab: distance-comparison} that the distances measured by ED are identical in all scenarios, except for Scenario~\#4, which partially overlaps with the ideal profile. This highlights ED's inability to capture temporal dynamics, as it lacks the sensitivity required to distinguish load shifts that align with the intended direction of the ideal profile. In contrast, DTW produces different distances for each scenario. However, in Scenario~\#1, the DTW value contradicts domain knowledge, as discussed in Section~\ref{sec:definition}.

In Scenario~\#3, both ED and DTW do not reflect the effectiveness of the rescheduling. Distances $M(\mathbf{O}, \mathbf{E})$ and $M(\mathbf{S_3}, \mathbf{E})$ are nearly identical for these metrics, even though Scenario~\#3 involves shifting a substantial portion of the evening peak load to noon. This adjustment more closely resembles the ideal load profile and therefore should be evaluated more favourably.

Scenario~\#2 emerges as the second-best option after the ideal profile, as it shifts the second-largest peak load to noon and results in a simplified single-peak structure. From both a pattern-matching and load-rescheduling perspective, this scenario is similar to the ideal profile. While FD correctly identifies Scenario~\#2 as the next most effective rescheduling option, ED and DTW fail to capture this relationship. 

This example demonstrates that, overall, FD provides the most interpretable and meaningful results when evaluating flexibility in load profiles, particularly in the context of time-aware rescheduling decisions.

 \begin{table}[!tb]
    \centering
    \scriptsize
    \caption{Distance values for all scenarios obtained by ED, DTW, and FD (brighter blue indicates a smaller distance value for each column)}
    \label{tab: distance-comparison}
    \renewcommand{\arraystretch}{1.5}
    \begin{tabular}{ccccc}
        \hline
        \rowcolor{TBLHeader}Distance & & ED & DTW & FD\\
\cline{1-1} \cline{3-5}
         $M(\mathbf{O}, \mathbf{E})$ & & \cellcolor{TBBRow6}1.45 & \cellcolor{TBBRow6}1.17 & \cellcolor{TBBRow6}1.44\\
         \hhline{|-|~|---|}
         $M(\mathbf{S_1}, \mathbf{E})$& & \cellcolor{TBBRow6}1.45 & \cellcolor{TBBRow4}1.07 & \cellcolor{TBBRow6}1.44 \\
         \hhline{|-|~|---|}
         $M(\mathbf{S_2}, \mathbf{E})$ & & \cellcolor{TBBRow6}1.45  & \cellcolor{TBBRow2}0.82 & \cellcolor{TBBRow1}0.46 \\
         \hhline{|-|~|---|}
         $M(\mathbf{S_3}, \mathbf{E})$ & & \cellcolor{TBBRow6}1.45 & \cellcolor{TBBRow5}1.16 & \cellcolor{TBBRow4}0.96 \\
         \hhline{|-|~|---|}
         $M(\mathbf{S_4}, \mathbf{E})$ & & \cellcolor{TBBRow1}1.04 & \cellcolor{TBBRow1}0.78 & \cellcolor{TBBRow2}0.62 \\
         \hhline{|-|~|---|}
         $M(\mathbf{S_5}, \mathbf{E})$ & & \cellcolor{TBBRow6}1.45 & \cellcolor{TBBRow3}0.89 & \cellcolor{TBBRow3}0.93 \\
         \hline
    \end{tabular}
\end{table}

\section{Conclusion and Future Works}
\label{sec:conclusions}

This paper proposed a novel distance metric, termed \emph{Flexibility Distance (FD)}, designed to quantify the effort required to reshape one time series into another. FD is particularly suitable for time series analysis involving dynamic user behaviour, such as electricity consumption data. Unlike traditional measures such as ED and DTW, the proposed FD metric jointly considers \emph{pointwise changes}, \emph{temporal shifts}, and \emph{sequence reordering}. This comprehensive consideration is especially critical in the context of indirect control in CDR, where the delivery of flexibility is not a binary outcome (i.e., fully delivered or not delivered at all). FD thus enables a more accurate quantification of how closely a realised consumption profile aligns with a targeted profile.

Moreover, given that the widely used mean squared error can be interpreted as a normalised form of the sum of squared ED values, FD may serve as a refined loss or error function in machine learning-based smart grid applications. By incorporating both amplitude and temporal differences, FD can guide learning algorithms more effectively, potentially leading to improved application performance, such as enhanced energy efficiency, reduced renewable energy curtailment, and greater customer satisfaction.

Future work will focus on extending the application of FD to a broader range of smart grid tasks, including load clustering, behind-the-meter (BTM) equipment identification, anomaly detection, load disaggregation, and data compression. In addition, our goal is to develop methods for generating \emph{temporal weights} that reflect user preferences or operational priorities. Finally, FD will be integrated into advanced learning frameworks such as contrastive learning and motif discovery, and evaluated using large-scale, high-resolution datasets. A comparative performance analysis with ED-based loss functions will also be performed.


\section*{Declaration of generative AI and AI-assisted technologies in the writing process}
During the preparation of this work the author(s) used ChatGPT and Grammarly to check grammar and improve phrasing. After using this tool/service, the author(s) reviewed and edited the content as needed and take(s) full responsibility for the content of the publication.

\balance


\bibliographystyle{IEEEtran}
\bibliography{reference}

\begin{thebibliography}{10}
\providecommand{\url}[1]{#1}
\csname url@samestyle\endcsname
\providecommand{\newblock}{\relax}
\providecommand{\bibinfo}[2]{#2}
\providecommand{\BIBentrySTDinterwordspacing}{\spaceskip=0pt\relax}
\providecommand{\BIBentryALTinterwordstretchfactor}{4}
\providecommand{\BIBentryALTinterwordspacing}{\spaceskip=\fontdimen2\font plus
\BIBentryALTinterwordstretchfactor\fontdimen3\font minus \fontdimen4\font\relax}
\providecommand{\BIBforeignlanguage}[2]{{%
\expandafter\ifx\csname l@#1\endcsname\relax
\typeout{** WARNING: IEEEtran.bst: No hyphenation pattern has been}%
\typeout{** loaded for the language `#1'. Using the pattern for}%
\typeout{** the default language instead.}%
\else
\language=\csname l@#1\endcsname
\fi
#2}}
\providecommand{\BIBdecl}{\relax}
\BIBdecl

\bibitem{IEA2023}
{International Energy Agency}, ``Efficient grid-interactive buildings,'' Tech. Rep. January, 2023.

\bibitem{GRANELL2015}
\BIBentryALTinterwordspacing
R.~Granell, C.~J. Axon, and D.~C. Wallom, ``{Clustering disaggregated load profiles using a Dirichlet process mixture model},'' \emph{Energy Conversion and Management}, vol.~92, pp. 507--516, 2015. [Online]. Available: \url{https://www.sciencedirect.com/science/article/pii/S0196890414011194}
\BIBentrySTDinterwordspacing

\bibitem{YUAN2023105588}
\BIBentryALTinterwordspacing
R.~Yuan, S.~A. Pourmousavi, W.~L. Soong, G.~Nguyen, and J.~A. Liisberg, ``Irmac: Interpretable refined motifs in binary classification for smart grid applications,'' \emph{Engineering Applications of Artificial Intelligence}, vol. 117, p. 105588, 2023. [Online]. Available: \url{https://www.sciencedirect.com/science/article/pii/S0952197622005784}
\BIBentrySTDinterwordspacing

\bibitem{Lin2012}
J.~Lin, R.~Khade, and Y.~Li, ``{Rotation-invariant similarity in time series using bag-of-patterns representation},'' \emph{Journal of Intelligent Information Systems}, vol.~39, no.~2, pp. 287--315, 2012.

\bibitem{Fang2018}
Z.~Fang, P.~Wang, and W.~Wang, ``{Efficient learning interpretable shapelets for accurate time series classification},'' \emph{Proceedings - IEEE 34th International Conference on Data Engineering, ICDE 2018}, pp. 497--508, 2018.

\bibitem{Nichiforov2021}
C.~Nichiforov and M.~Alamaniotis, ``Load-based classification of academic buildings using matrix profile and supervised learning,'' pp. 01--05, 2021.

\bibitem{shaw2025forestproximitiestimeseries}
\BIBentryALTinterwordspacing
B.~Shaw, J.~Rhodes, S.~F. Boubrahimi, and K.~R. Moon, ``Forest proximities for time series,'' 2025. [Online]. Available: \url{https://arxiv.org/abs/2410.03098}
\BIBentrySTDinterwordspacing

\bibitem{Elafoudi2014}
G.~Elafoudi, L.~Stankovic, and V.~Stankovic, ``Power disaggregation of domestic smart meter readings using dynamic time warping,'' in \emph{2014 6th International Symposium on Communications, Control and Signal Processing (ISCCSP)}, 2014, pp. 36--39.

\bibitem{Wen2023}
Y.~Wen, Z.~Hu, J.~He, and Y.~Guo, ``Improved inner approximation for aggregating power flexibility in active distribution networks and its applications,'' \emph{IEEE Transactions on Smart Grid}, vol.~15, no.~4, pp. 3653--3665, 2023.

\bibitem{Li2024}
Y.~Li and Z.~Li, ``Distributionally robust evaluation for real-time flexibility of electric vehicles considering uncertain departure behavior and state-of-charge,'' \emph{IEEE Transactions on Smart Grid}, vol.~15, no.~4, pp. 4288--4291, 2024.

\bibitem{Goldenberg2018}
C.~Goldenberg, M.~Dyson, and H.~Masters, ``Demand flexibility the key to enabling a low-cost, low-carbon grid,'' no. February, 2018.

\bibitem{Gerards2017}
M.~E.~T. Gerards and J.~L. Hurink, ``On the value of device flexibility in smart grid applications,'' in \emph{2017 IEEE Manchester PowerTech}, 2017, pp. 1--6.

\bibitem{Varenhorst2024}
I.~A.~M. Varenhorst, M.~E.~T. Gerards, and J.~L. Hurink, ``Quantifying device flexibility with shapley values in demand side management,'' in \emph{e-Energy '24}, 2024, pp. 147--157.

\bibitem{flex2012}
E.~Lannoye, D.~Flynn, and M.~O'Malley, ``Evaluation of power system flexibility,'' \emph{IEEE Transactions on Power Systems}, vol.~27, no.~2, pp. 922--931, 2012.

\bibitem{Keogh2005}
E.~Keogh and C.~A. Ratanamahatana, ``{Exact indexing of dynamic time warping},'' \emph{Knowledge and Information Systems}, vol.~7, no.~3, pp. 358--386, 2005.

\bibitem{Bougleux2020}
S.~Bougleux, B.~Ga{\"{u}}z{\`{e}}re, D.~B. Blumenthal, and L.~Brun, ``{Fast linear sum assignment with error-correction and no cost constraints},'' \emph{Pattern Recognition Letters}, vol. 134, pp. 37--45, 2020.

\bibitem{datasourcewebsite}
\BIBentryALTinterwordspacing
{Ausgrid}. (2015) Solar home electricity data. [Online]. Available: \url{"https://www.ausgrid.com.au/Industry/Our-Research/Data-to-share/Solar-home-electricity-data"}
\BIBentrySTDinterwordspacing

\end{thebibliography}

\vfill

\end{document}